\documentstyle[aps,pre,multicol]{revtex}
\begin{document}
\draft
\title{The critical Ising lines of the $d=2$ Ashkin-Teller model}
\author{G. Kamieniarz and P. Koz{\l}owski}
\address{Computational Physics Division Institute of Physics,
A.Mickiewicz University,\\
ul. Umultowska 85, PL 61-624 Pozna\'{n}, Poland\\
{\em e-mail:} {\tt gjk@pearl.amu.edu.pl}}
\author{R. Dekeyser}
\address{Institute for Theoretical Physics, Katholieke Universiteit
Leuven,\\ Celestijnenlaan 200D, B 3001 Leuven, Belgium\\
{\em e-mail:} {\tt raf.dekeyser@fys.kuleuven.ac.be}}
\date{\today}           
\maketitle
\begin{abstract}
The universal critical point ratio $Q$ is exploited to determine
positions of the critical Ising transition lines on the phase diagram of the
Ashkin-Teller (AT) model on the square lattice. A leading-order expansion of
the ratio $Q$ in the presence of a non-vanishing thermal field is
found from finite-size scaling and the corresponding expression is
fitted to the accurate perturbative transfer-matrix data calculations
for the $L\times L$ square clusters with $L\leq 9$.
\end{abstract}
\begin{multicols}{2}
\narrowtext

The AT model has first been proposed as a model of a four component alloy
\cite{AT}. It has attracted a lot of theoretical interest for years because
it is a simple and non-trivial generalization of the Ising and four-state
Potts models. Fan \cite{Fan1} has shown that the hamiltonian of the AT model
can also be written  with
two Ising variables ($S=\pm 1, \sigma=\pm 1$) located at each site of the
lattice, which in the presence of a magnetic field has the form:

\begin{eqnarray}
{\cal H}=-\sum_{<i,j>}(J_{1}S_{i}S_{j} + J_{2}\sigma_{i}\sigma_{j}\nonumber \\
+J_{4}S_{i}\sigma_{i}S_{j}\sigma_{j} + J_{0})
-h \sum_{i=1}^{N}S_{i}\sigma_{i}  \label{ham}
\end{eqnarray}

\noindent Herein we consider only the nearest neighbour pair interactions  on
the simple square lattice consisting of $N=L^{2}$ sites with periodic
boundary conditions
and we assume that $J_{1}=J_{2}$ (isotropic case).

Wagner \cite{Wagn} has shown that the AT model is equivalent to the alternated
eight vertex model, which has not been solved exactly.
Only one critical line in the phase diagram of the isotropic AT model
is known exactly thanks to the duality relation found by Fan \cite{Fan2}.
For this reason many approximate approaches have been applied for
constructing the
complete phase diagram: the mean field theory (MFA)
\cite{Dit_al,Paw-Rog},
mean-field renormalisation group (MFRG) \cite{Pla_Ba},
renormalisation group (RG)  \cite{Ben_al}, and Monte Carlo renormalisation
group (MCRG)  \cite{Cha_al}.
It is the aim of this paper
to establish an accurate location of the remaining critical
lines.

In our approach we exploit finite-size scaling for the ratio of the square of
the second moment to the fourth moment of the order parameter $M$:

\begin{eqnarray}
Q_{L}=\frac{<M^{2}>^{2}_{L}}{<M^{4}>_{L}}~, \label{ql}
\end{eqnarray}

\noindent where $<...>$ means thermal average
and the index
$L$ indicates the linear size of the system ($L\times L$).
In the limit $L\rightarrow \infty$ this ratio becomes
universal in the critical point
\cite{priv} and is denoted $Q$ hereafter.
 Three not exactly known critical lines of
the isotropic AT model
 are believed to belong to the Ising
universality class \cite{Dit_al,Bax}.
Here it is assumed that these lines correspond
to the Ising-like continuous transitions with the order parameter
$M=\sum_{i=1}^{N}S_{i}\sigma_{i}$.
A scaling formula for $Q_{L}$ can be derived starting from
the finite-size scaling relation for the singular part of the free energy for
the square Ising model
\cite{Kam-Blot}.

\begin{equation}
F^{(S)}(g_{t},g_{h},L^{-1})=A(g_{t}L)\ln L+B(g_{t}L,g_{h}L^{y_{h}}) \label{fs}
\end{equation}

\noindent where $A$ and $B$ are unknown amplitudes, $g_{t}$, $g_{h}$ are
nonlinear scaling fields and $y_{h}$ is the magnetic critical exponent. The
nonlinear scaling fields $g_{t}$ and $g_{h}$ can be expanded in terms of the
corresponding linear thermal and magnetic scaling fields $t$ and $h$.




Taking into account the relations between the magnetization moments
in Eq.\ (\ref{ql}) and
the corresponding
derivatives of the free energy \cite{Kam-Blot}
\noindent we have calculated the scaling expansion for $Q_{L}(t,h=0)$ to the
leading order in $t$ and up to $L^{3-4y_{h}}$:

\begin{equation}
Q_{L}(t)=Q_{L}(0)+\left.\frac{\partial Q_{L}(t)}{\partial t}\right|_{t=0}t+...
\label{qt}
\end{equation}

\noindent
The zeroth order term $Q_{L}(0)$ was evaluated previously
\cite{Kam-Blot} and the first order term is of the form
\end{multicols}
\widetext

\noindent\rule{20.5pc}{0.1mm}\rule{0.1mm}{1.5mm}\hfill
\begin{eqnarray}
\label{dqlt}
\left.\frac{\partial Q_{L}(t)}{\partial t}\right|_{t=0}=\alpha_{1}L+\alpha_{2}+
\alpha_{3}L^{3-2y_{h}}+(\alpha_{4}+\alpha_{5}\ln L)L^{2-2y_{h}}
+\alpha_{6}L^{5-4y_{h}}+\nonumber\\
+(\alpha_{7}+\alpha_{8}\ln L)L^{1-2y_{h}}+
(\alpha_{9}+\alpha_{10}\ln L)L^{4-4y_{h}}
+\alpha_{11}L^{-2y_{h}}+\\
+(\alpha_{12}+\alpha_{13}\ln L)L^{7-6y_{h}}
+(\alpha_{14}+\alpha_{15}\ln L+ \alpha_{16}\ln^{2}L)L^{3-4y_{h}}+
...~,\nonumber
\end{eqnarray}

\hfill\rule[-1.5mm]{0.1mm}{1.5mm}\rule{20.5pc}{0.1mm}
\begin{multicols}{2}
\narrowtext
\noindent where $\alpha_{i}$ ($i=1,...16$) are unknown amplitudes.
In our work we consider only the first three terms in the expansion
(\ref{dqlt}),
but for some future Monte Carlo applications the higher order terms in $1/L$
might be important.

We have calculated the $Q_{L}(t)$ ratio exploiting the transfer matrix technique
which for the Ising model was explained in \cite{Kam-Blot}. Our system consists
of $L$ columns containing $L$ sites.
Spins from the $j$th column are denoted by
$\vec{\Sigma}_{j}=(S_{j1},\sigma_{j1},S_{j2},\sigma_{j2},...,
S_{jL},\sigma_{jL})$ so that

\begin{equation}
Z=\sum_{\vec{\Sigma}_{1},\vec{\Sigma}_{2},...,\vec{\Sigma}_{L}}
\exp (-\beta {\it H}(\vec{\Sigma}_{1},...,\vec{\Sigma}_{L}))=
{\rm Tr}\, {\bf T}^{L}~,  \label{ztr}
\end{equation}

\noindent where {\bf T} is a $4^{L}\times 4^{L}$ transfer matrix. This can be
split into the product ${\bf T}={\bf T}_{h}{\bf T}_{v}$ of a diagonal
matrix ${\bf T}_{v}$
and a non-diagonal matrix ${\bf T}_{h}$ containing the intra- and the
inter-column interactions respectively.
They are defined as follows
\end{multicols}
\widetext

\noindent\rule{20.5pc}{0.1mm}\rule{0.1mm}{1.5mm}\hfill
\begin{equation}
{\bf T}_{v}(\vec{\Sigma}_{k},\vec{\Sigma}_{l})=
\delta_{\vec{\Sigma}_{k},\vec{\Sigma}_{l}}\exp \left(\sum_{i=1}^{N}
(K_{2}S_{k,i}S_{k,i+1}+K_{2}\sigma_{k,i}\sigma_{k,i+1}
+K_{4}S_{k,i}\sigma_{k,i}S_{k,i+1}\sigma_{k,i+1}+HS_{k,i}
\sigma_{k,i})\right)
\end{equation}
\begin{equation}
{\bf T}_{h}(\vec{\Sigma}_{k},\vec{\Sigma}_{l})=
\exp \left(\sum_{i=1}^{N}(K_{2}S_{k,i}S_{l,i}+K_{2}\sigma_{k,i}\sigma_{l,i}+
K_{4}S_{k,i}\sigma_{k,i}S_{l,i}\sigma_{l,i})\right),
\end{equation}
\hfill\rule[-1.5mm]{0.1mm}{1.5mm}\rule{20.5pc}{0.1mm}
\begin{multicols}{2}
\narrowtext
\noindent where
, $\beta=\frac{1}{k_{B}T}$,
$K_{i}=J_{i}\beta$ ($i=1,2,4$) and
$H=\beta h$. The latter matrix can be expressed as a product of
sparse matrices which facilitates the numerical calculations.

The averages in Eq.\ (\ref{ql}) can be expressed in terms of
the corresponding coefficients $Z_{k}$ \cite{Kam-Blot} in the expansion of
the field dependent partition function
$Z(h)=\sum_{k=0}^{\infty}Z_{k}\frac{h^{k}}{k!}$.
%
%
The coefficients $Z_{k}$ can then be calculated from Eq.\ (\ref{ztr})
by
multiplying the base vectors by matrices ${\bf T}_{v}$ and ${\bf T}_{h}$
in such a manner that the terms in the same power of $h$ are kept
separately \cite{Kam-Blot}.

At first we calculate the amplitudes $\alpha_{i}$ ($i\leq 5$) from Eqs
(\ref{qt}) and (\ref{dqlt}) with known values $Q_{L}(0)$. In the limit
$K_{2}=0$,
i.e. the Ising model in $S\sigma$, $K_{4c}=K_{c}=\frac{1}{2}\ln (1+\sqrt{2})$
and in this case we have only one coupling constant ($K_{4}$). Thus we can write
the reduced temperature in the form:

\begin{equation}
t=\frac{K_{4c}-K_{4}}{K_{4}}~. \label{t}
\end{equation}

Selecting different values of the scaling field $t$ we can solve the set of
linear algebraic equations for $\alpha_{i}$. For the ferromagnetic coupling
$K_{4}$ we consider the system sizes $L=2,3,...,9$ whereas for the
anti-ferromagnetic one only the even values $L=2,4,6,8$ are considered, so that
we can evaluate
the coefficients $\alpha_{i}$ up to $i=5$ or $i=3$, respectively.

Having fixed $K_{2}\neq 0$ and knowing the $\alpha_{i}$ ($i\leq 3$) and
$Q_{L}(0)$, we have calculated $Q_{L}(K_{2},K_{4})$ for a number of couplings
$K_{4}$. This enables a determination of the corresponding $t$ values from Eqs
(\ref{qt}) and
(\ref{dqlt}). Then knowing $t$ we can easily obtain $K_{4c}$ from Eq.\ (\ref{t})
 and $K_{2c}$ from a similar equation, but written for $K_{2}$.
%
%
 The estimates $K_{4c}$ and $K_{2c}$ are
very stable if we find $t\in <10^{-7}, 10^{-4}>$.

The exactly known critical curve  with continuously varying critical exponents
\cite{Bax} is terminated in the 4-state Potts point where it bifurcates. In the
vicinity of this point the convergence of our results is diminished and the
estimates of $K_{4c}$ become size dependent. This size dependence is illustrated
in Fig 1. Due to the limited number of system sizes available in our
calculations we
do not try to include any corrections to scaling and we simply extrapolate our
data. The corresponding estimates are shown on the ordinate axis in Fig 1.
Such a strong size dependence does not occur for the anti-ferromagnetic
couplings,
since there is no Potts point in this case.

Our final results represented by open circles connected by
thin continuous lines are shown in Fig.~\ref{res}
 and they are compared with other results and
predictions. The numerical uncertainties do not exceed the
size
of the symbol.
The curve plotted by the bold line represents the part of the phase diagram
found exactly by Baxter \cite{Bax}. It separates the Baxter phase B from the
paramagnetic phase P. The ferromagnetic and anti-ferromagnetic phases with
non-vanishing order parameter $M$ are denoted by the labels F
and AF,
respectively.

In the ferromagnetic region $K_{4} > 0$ we have only calculated the curve
joining
the 4-state Potts point to the pure Ising point $K_{c}$ at $K_{2}=0$. The second
branch follows from the corresponding duality relation \cite{Dit_al,Bax}.
In the boundary between  AF and P phases with
the dotted lines we plot the approximate curve as given by Baxter
\cite{Bax}  and in the ferromagnetic region
we also include the MCRG results marked by filled circles.

As can be seen (Fig.~\ref{res}) our results are in good
agreement with the MCRG \cite{Cha_al} approach, but are quite different
 from Baxter's predictions \cite{Bax} in the anti-ferromagnetic region.
 For the boundary between AF and P phases, our results
 coincide with those
 obtained by Mazzeo {\it et al.} \cite{enri}. These authors actually
 investigated the six vertex model with the transfer matrix technique in
 combination with conformal invariance arguments; their results can be
 mapped onto the results for the P-phase boundaries and they are shown in
 Fig.\ 2.

As to our accuracy: near the ferromagnetic Ising point
 it is of about $2*10^{-6}$
and in the neighbourhood of the Potts point
it decreases down to about $3*10^{-2}$.
The accuracy in the anti-ferromagnetic region is even better:
near the Ising point it reaches $5*10^{-8}$ and
for the highest point at the phase diagram in Fig.~\ref{res} it decreases to
$3*10^{-3}$.

The numerical calculations were carried out in the Supercomputing and
Networking Center in Pozna\'{n} on Cray J-916. The work has been
supported in part by the Committee for the Scientific Research via grant 2 P302
116 06. We thank also Dr. E. Carlon, Dr. P. Pawlicki
and Prof. J. Rogiers for discussions.

\end{multicols} \widetext

\begin{figure}[h]
\begin{center}

\setlength{\unitlength}{0.240900pt}
\ifx\plotpoint\undefined\newsavebox{\plotpoint}\fi
\sbox{\plotpoint}{\rule[-0.200pt]{0.400pt}{0.400pt}}%
\begin{picture}(1875,900)(0,0)
\font\gnuplot=cmr10 at 10pt
\gnuplot
\sbox{\plotpoint}{\rule[-0.200pt]{0.400pt}{0.400pt}}%
\put(220.0,113.0){\rule[-0.200pt]{0.400pt}{184.048pt}}
\put(220.0,192.0){\rule[-0.200pt]{4.818pt}{0.400pt}}
\put(198,192){\makebox(0,0)[r]{$0.32$}}
\put(1791.0,192.0){\rule[-0.200pt]{4.818pt}{0.400pt}}
\put(220.0,306.0){\rule[-0.200pt]{4.818pt}{0.400pt}}
\put(198,306){\makebox(0,0)[r]{$0.34$}}
\put(1791.0,306.0){\rule[-0.200pt]{4.818pt}{0.400pt}}
\put(220.0,419.0){\rule[-0.200pt]{4.818pt}{0.400pt}}
\put(198,419){\makebox(0,0)[r]{$0.36$}}
\put(1791.0,419.0){\rule[-0.200pt]{4.818pt}{0.400pt}}
\put(220.0,533.0){\rule[-0.200pt]{4.818pt}{0.400pt}}
\put(198,533){\makebox(0,0)[r]{$0.38$}}
\put(1791.0,533.0){\rule[-0.200pt]{4.818pt}{0.400pt}}
\put(220.0,646.0){\rule[-0.200pt]{4.818pt}{0.400pt}}
\put(198,646){\makebox(0,0)[r]{$0.4$}}
\put(1791.0,646.0){\rule[-0.200pt]{4.818pt}{0.400pt}}
\put(220.0,760.0){\rule[-0.200pt]{4.818pt}{0.400pt}}
\put(198,760){\makebox(0,0)[r]{$0.42$}}
\put(1791.0,760.0){\rule[-0.200pt]{4.818pt}{0.400pt}}
\put(220.0,873.0){\rule[-0.200pt]{4.818pt}{0.400pt}}
\put(198,873){\makebox(0,0)[r]{$0.44$}}
\put(1791.0,873.0){\rule[-0.200pt]{4.818pt}{0.400pt}}
\put(220.0,113.0){\rule[-0.200pt]{0.400pt}{4.818pt}}
\put(220,68){\makebox(0,0){$0$}}
\put(220.0,857.0){\rule[-0.200pt]{0.400pt}{4.818pt}}
\put(509.0,113.0){\rule[-0.200pt]{0.400pt}{4.818pt}}
\put(509,68){\makebox(0,0){$0.1$}}
\put(509.0,857.0){\rule[-0.200pt]{0.400pt}{4.818pt}}
\put(799.0,113.0){\rule[-0.200pt]{0.400pt}{4.818pt}}
\put(799,68){\makebox(0,0){$0.2$}}
\put(799.0,857.0){\rule[-0.200pt]{0.400pt}{4.818pt}}
\put(1088.0,113.0){\rule[-0.200pt]{0.400pt}{4.818pt}}
\put(1088,68){\makebox(0,0){$0.3$}}
\put(1088.0,857.0){\rule[-0.200pt]{0.400pt}{4.818pt}}
\put(1377.0,113.0){\rule[-0.200pt]{0.400pt}{4.818pt}}
\put(1377,68){\makebox(0,0){$0.4$}}
\put(1377.0,857.0){\rule[-0.200pt]{0.400pt}{4.818pt}}
\put(1666.0,113.0){\rule[-0.200pt]{0.400pt}{4.818pt}}
\put(1666,68){\makebox(0,0){$0.5$}}
\put(1666.0,857.0){\rule[-0.200pt]{0.400pt}{4.818pt}}
\put(220.0,113.0){\rule[-0.200pt]{383.272pt}{0.400pt}}
\put(1811.0,113.0){\rule[-0.200pt]{0.400pt}{184.048pt}}
\put(220.0,877.0){\rule[-0.200pt]{383.272pt}{0.400pt}}
\put(45,495){\makebox(0,0){$K_{4c}$}}
\put(1015,23){\makebox(0,0){$L^{-1}$}}
\put(220.0,113.0){\rule[-0.200pt]{0.400pt}{184.048pt}}
\put(220,873){\raisebox{-.8pt}{\makebox(0,0){$\Diamond$}}}
\put(582,873){\raisebox{-.8pt}{\makebox(0,0){$\Diamond$}}}
\put(633,873){\raisebox{-.8pt}{\makebox(0,0){$\Diamond$}}}
\put(702,873){\raisebox{-.8pt}{\makebox(0,0){$\Diamond$}}}
\put(799,873){\raisebox{-.8pt}{\makebox(0,0){$\Diamond$}}}
\put(943,873){\raisebox{-.8pt}{\makebox(0,0){$\Diamond$}}}
\put(1184,872){\raisebox{-.8pt}{\makebox(0,0){$\Diamond$}}}
\put(1666,868){\raisebox{-.8pt}{\makebox(0,0){$\Diamond$}}}
\put(220,859){\makebox(0,0){$+$}}
\put(582,859){\makebox(0,0){$+$}}
\put(633,859){\makebox(0,0){$+$}}
\put(702,859){\makebox(0,0){$+$}}
\put(799,859){\makebox(0,0){$+$}}
\put(943,859){\makebox(0,0){$+$}}
\put(1184,857){\makebox(0,0){$+$}}
\put(1666,842){\makebox(0,0){$+$}}
\sbox{\plotpoint}{\rule[-0.400pt]{0.800pt}{0.800pt}}%
\put(220,836){\raisebox{-.8pt}{\makebox(0,0){$\Box$}}}
\put(582,836){\raisebox{-.8pt}{\makebox(0,0){$\Box$}}}
\put(633,836){\raisebox{-.8pt}{\makebox(0,0){$\Box$}}}
\put(702,836){\raisebox{-.8pt}{\makebox(0,0){$\Box$}}}
\put(799,836){\raisebox{-.8pt}{\makebox(0,0){$\Box$}}}
\put(943,835){\raisebox{-.8pt}{\makebox(0,0){$\Box$}}}
\put(1184,829){\raisebox{-.8pt}{\makebox(0,0){$\Box$}}}
\put(1666,797){\raisebox{-.8pt}{\makebox(0,0){$\Box$}}}
\sbox{\plotpoint}{\rule[-0.500pt]{1.000pt}{1.000pt}}%
\put(220,802){\makebox(0,0){$\times$}}
\put(582,802){\makebox(0,0){$\times$}}
\put(633,802){\makebox(0,0){$\times$}}
\put(702,802){\makebox(0,0){$\times$}}
\put(799,801){\makebox(0,0){$\times$}}
\put(943,797){\makebox(0,0){$\times$}}
\put(1184,782){\makebox(0,0){$\times$}}
\put(1666,732){\makebox(0,0){$\times$}}
\sbox{\plotpoint}{\rule[-0.600pt]{1.200pt}{1.200pt}}%
\put(220,754){\makebox(0,0){$\triangle$}}
\put(541,753){\makebox(0,0){$\triangle$}}
\put(582,753){\makebox(0,0){$\triangle$}}
\put(633,753){\makebox(0,0){$\triangle$}}
\put(702,752){\makebox(0,0){$\triangle$}}
\put(799,748){\makebox(0,0){$\triangle$}}
\put(943,738){\makebox(0,0){$\triangle$}}
\put(1184,710){\makebox(0,0){$\triangle$}}
\put(1666,648){\makebox(0,0){$\triangle$}}
\sbox{\plotpoint}{\rule[-0.500pt]{1.000pt}{1.000pt}}%
\put(220,689){\makebox(0,0){$\star$}}
\put(541,685){\makebox(0,0){$\star$}}
\put(582,684){\makebox(0,0){$\star$}}
\put(633,682){\makebox(0,0){$\star$}}
\put(702,678){\makebox(0,0){$\star$}}
\put(799,668){\makebox(0,0){$\star$}}
\put(943,648){\makebox(0,0){$\star$}}
\put(1184,610){\makebox(0,0){$\star$}}
\put(1666,546){\makebox(0,0){$\star$}}
\sbox{\plotpoint}{\rule[-0.200pt]{0.400pt}{0.400pt}}%
\put(220,601){\circle{12}}
\put(541,586){\circle{12}}
\put(582,582){\circle{12}}
\put(633,575){\circle{12}}
\put(702,565){\circle{12}}
\put(799,548){\circle{12}}
\put(943,522){\circle{12}}
\put(1184,482){\circle{12}}
\put(1666,428){\circle{12}}
\put(220,459){\circle{18}}
\put(541,427){\circle{18}}
\put(582,419){\circle{18}}
\put(633,409){\circle{18}}
\put(702,396){\circle{18}}
\put(799,379){\circle{18}}
\put(943,356){\circle{18}}
\put(1184,325){\circle{18}}
\put(1666,288){\circle{18}}
\sbox{\plotpoint}{\rule[-0.400pt]{0.800pt}{0.800pt}}%
\put(220,221){\circle{24}}
\put(541,197){\circle{24}}
\put(582,191){\circle{24}}
\put(633,184){\circle{24}}
\put(702,176){\circle{24}}
\put(799,165){\circle{24}}
\put(943,152){\circle{24}}
\put(1184,134){\circle{24}}
\put(1666,116){\circle{24}}
\end{picture}

\end{center}
\caption{The L-dependence of the critical values of the parameter $K_{4c}$. The
points on the
vertical axis are the extrapolated values.} \label{k4l}
\end{figure}
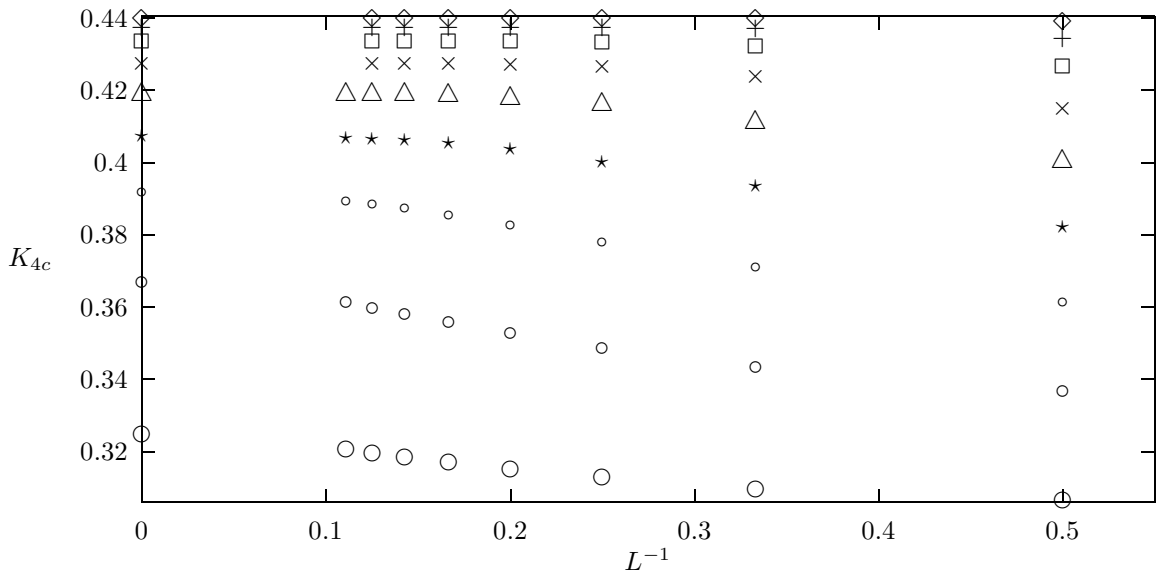

\begin{figure}[h]
\begin{center}
\setlength{\unitlength}{0.240900pt}
\ifx\plotpoint\undefined\newsavebox{\plotpoint}\fi
\sbox{\plotpoint}{\rule[-0.200pt]{0.400pt}{0.400pt}}%
\begin{picture}(1875,1034)(0,0)
\font\gnuplot=cmr10 at 10pt
\gnuplot
\sbox{\plotpoint}{\rule[-0.200pt]{0.400pt}{0.400pt}}%
\put(220.0,113.0){\rule[-0.200pt]{4.818pt}{0.400pt}}
\put(198,113){\makebox(0,0)[r]{$0$}}
\put(1791.0,113.0){\rule[-0.200pt]{4.818pt}{0.400pt}}
\put(220.0,225.0){\rule[-0.200pt]{4.818pt}{0.400pt}}
\put(198,225){\makebox(0,0)[r]{$0.1$}}
\put(1791.0,225.0){\rule[-0.200pt]{4.818pt}{0.400pt}}
\put(220.0,337.0){\rule[-0.200pt]{4.818pt}{0.400pt}}
\put(198,337){\makebox(0,0)[r]{$0.2$}}
\put(1791.0,337.0){\rule[-0.200pt]{4.818pt}{0.400pt}}
\put(220.0,449.0){\rule[-0.200pt]{4.818pt}{0.400pt}}
\put(198,449){\makebox(0,0)[r]{$0.3$}}
\put(1791.0,449.0){\rule[-0.200pt]{4.818pt}{0.400pt}}
\put(220.0,561.0){\rule[-0.200pt]{4.818pt}{0.400pt}}
\put(198,561){\makebox(0,0)[r]{$0.4$}}
\put(1791.0,561.0){\rule[-0.200pt]{4.818pt}{0.400pt}}
\put(220.0,673.0){\rule[-0.200pt]{4.818pt}{0.400pt}}
\put(198,673){\makebox(0,0)[r]{$0.5$}}
\put(1791.0,673.0){\rule[-0.200pt]{4.818pt}{0.400pt}}
\put(220.0,785.0){\rule[-0.200pt]{4.818pt}{0.400pt}}
\put(198,785){\makebox(0,0)[r]{$0.6$}}
\put(1791.0,785.0){\rule[-0.200pt]{4.818pt}{0.400pt}}
\put(220.0,897.0){\rule[-0.200pt]{4.818pt}{0.400pt}}
\put(198,897){\makebox(0,0)[r]{$0.7$}}
\put(1791.0,897.0){\rule[-0.200pt]{4.818pt}{0.400pt}}
\put(220.0,1009.0){\rule[-0.200pt]{4.818pt}{0.400pt}}
\put(198,1009){\makebox(0,0)[r]{$0.8$}}
\put(1791.0,1009.0){\rule[-0.200pt]{4.818pt}{0.400pt}}
\put(365.0,113.0){\rule[-0.200pt]{0.400pt}{4.818pt}}
\put(365,68){\makebox(0,0){$-0.6$}}
\put(365.0,991.0){\rule[-0.200pt]{0.400pt}{4.818pt}}
\put(571.0,113.0){\rule[-0.200pt]{0.400pt}{4.818pt}}
\put(571,68){\makebox(0,0){$-0.4$}}
\put(571.0,991.0){\rule[-0.200pt]{0.400pt}{4.818pt}}
\put(778.0,113.0){\rule[-0.200pt]{0.400pt}{4.818pt}}
\put(778,68){\makebox(0,0){$-0.2$}}
\put(778.0,991.0){\rule[-0.200pt]{0.400pt}{4.818pt}}
\put(985.0,113.0){\rule[-0.200pt]{0.400pt}{4.818pt}}
\put(985,68){\makebox(0,0){$0$}}
\put(985.0,991.0){\rule[-0.200pt]{0.400pt}{4.818pt}}
\put(1191.0,113.0){\rule[-0.200pt]{0.400pt}{4.818pt}}
\put(1191,68){\makebox(0,0){$0.2$}}
\put(1191.0,991.0){\rule[-0.200pt]{0.400pt}{4.818pt}}
\put(1398.0,113.0){\rule[-0.200pt]{0.400pt}{4.818pt}}
\put(1398,68){\makebox(0,0){$0.4$}}
\put(1398.0,991.0){\rule[-0.200pt]{0.400pt}{4.818pt}}
\put(1604.0,113.0){\rule[-0.200pt]{0.400pt}{4.818pt}}
\put(1604,68){\makebox(0,0){$0.6$}}
\put(1604.0,991.0){\rule[-0.200pt]{0.400pt}{4.818pt}}
\put(1811.0,113.0){\rule[-0.200pt]{0.400pt}{4.818pt}}
\put(1811,68){\makebox(0,0){$0.8$}}
\put(1811.0,991.0){\rule[-0.200pt]{0.400pt}{4.818pt}}
\put(220.0,113.0){\rule[-0.200pt]{383.272pt}{0.400pt}}
\put(1811.0,113.0){\rule[-0.200pt]{0.400pt}{216.328pt}}
\put(220.0,1011.0){\rule[-0.200pt]{383.272pt}{0.400pt}}
\put(45,562){\makebox(0,0){$K_{2}$}}
\put(1015,23){\makebox(0,0){$K_{4}$}}
\put(1191,729){\makebox(0,0)[l]{B}}
\put(799,348){\makebox(0,0)[l]{P}}
\put(1584,259){\makebox(0,0)[l]{F}}
\put(365,259){\makebox(0,0)[l]{AF}}
\put(1475,141){\makebox(0,0)[l]{$K_{c}$}}
\put(416,141){\makebox(0,0)[l]{$-K_{c}$}}
\put(1191,538){\makebox(0,0)[l]{$q=4$ Potts}}
\put(1191,729){\makebox(0,0)[l]{B}}
\put(799,348){\makebox(0,0)[l]{P}}
\put(1584,259){\makebox(0,0)[l]{F}}
\put(365,259){\makebox(0,0)[l]{AF}}
\put(1475,141){\makebox(0,0)[l]{$K_{c}$}}
\put(416,141){\makebox(0,0)[l]{$-K_{c}$}}
\put(1191,538){\makebox(0,0)[l]{$q=4$ Potts}}
\put(1191,729){\makebox(0,0)[l]{B}}
\put(799,348){\makebox(0,0)[l]{P}}
\put(1584,259){\makebox(0,0)[l]{F}}
\put(365,259){\makebox(0,0)[l]{AF}}
\put(1475,141){\makebox(0,0)[l]{$K_{c}$}}
\put(416,141){\makebox(0,0)[l]{$-K_{c}$}}
\put(1191,538){\makebox(0,0)[l]{$q=4$ Potts}}
\put(1191,729){\makebox(0,0)[l]{B}}
\put(799,348){\makebox(0,0)[l]{P}}
\put(1584,259){\makebox(0,0)[l]{F}}
\put(365,259){\makebox(0,0)[l]{AF}}
\put(1475,141){\makebox(0,0)[l]{$K_{c}$}}
\put(416,141){\makebox(0,0)[l]{$-K_{c}$}}
\put(1191,538){\makebox(0,0)[l]{$q=4$ Potts}}
\put(1191,729){\makebox(0,0)[l]{B}}
\put(799,348){\makebox(0,0)[l]{P}}
\put(1584,259){\makebox(0,0)[l]{F}}
\put(365,259){\makebox(0,0)[l]{AF}}
\put(1475,141){\makebox(0,0)[l]{$K_{c}$}}
\put(416,141){\makebox(0,0)[l]{$-K_{c}$}}
\put(1191,538){\makebox(0,0)[l]{$q=4$ Potts}}
\put(1191,729){\makebox(0,0)[l]{B}}
\put(799,348){\makebox(0,0)[l]{P}}
\put(1584,259){\makebox(0,0)[l]{F}}
\put(365,259){\makebox(0,0)[l]{AF}}
\put(1475,141){\makebox(0,0)[l]{$K_{c}$}}
\put(416,141){\makebox(0,0)[l]{$-K_{c}$}}
\put(1191,538){\makebox(0,0)[l]{$q=4$ Potts}}
\put(220.0,113.0){\rule[-0.200pt]{0.400pt}{216.328pt}}
\multiput(1292.92,499.16)(-0.497,-1.649){49}{\rule{0.120pt}{1.408pt}}
\multiput(1293.17,502.08)(-26.000,-82.078){2}{\rule{0.400pt}{0.704pt}}
\put(1268,420){\vector(-1,-3){0}}
\multiput(1292.92,499.16)(-0.497,-1.649){49}{\rule{0.120pt}{1.408pt}}
\multiput(1293.17,502.08)(-26.000,-82.078){2}{\rule{0.400pt}{0.704pt}}
\put(1268,420){\vector(-1,-3){0}}
\multiput(1292.92,499.16)(-0.497,-1.649){49}{\rule{0.120pt}{1.408pt}}
\multiput(1293.17,502.08)(-26.000,-82.078){2}{\rule{0.400pt}{0.704pt}}
\put(1268,420){\vector(-1,-3){0}}
\multiput(1292.92,499.16)(-0.497,-1.649){49}{\rule{0.120pt}{1.408pt}}
\multiput(1293.17,502.08)(-26.000,-82.078){2}{\rule{0.400pt}{0.704pt}}
\put(1268,420){\vector(-1,-3){0}}
\multiput(1292.92,499.16)(-0.497,-1.649){49}{\rule{0.120pt}{1.408pt}}
\multiput(1293.17,502.08)(-26.000,-82.078){2}{\rule{0.400pt}{0.704pt}}
\put(1268,420){\vector(-1,-3){0}}
\multiput(1292.92,499.16)(-0.497,-1.649){49}{\rule{0.120pt}{1.408pt}}
\multiput(1293.17,502.08)(-26.000,-82.078){2}{\rule{0.400pt}{0.704pt}}
\put(1268,420){\vector(-1,-3){0}}
\put(1440,113){\usebox{\plotpoint}}
\put(1438.67,113){\rule{0.400pt}{7.468pt}}
\multiput(1439.17,113.00)(-1.000,15.500){2}{\rule{0.400pt}{3.734pt}}
\put(1437.17,144){\rule{0.400pt}{6.300pt}}
\multiput(1438.17,144.00)(-2.000,17.924){2}{\rule{0.400pt}{3.150pt}}
\multiput(1435.93,175.00)(-0.477,3.270){7}{\rule{0.115pt}{2.500pt}}
\multiput(1436.17,175.00)(-5.000,24.811){2}{\rule{0.400pt}{1.250pt}}
\multiput(1430.93,205.00)(-0.482,2.751){9}{\rule{0.116pt}{2.167pt}}
\multiput(1431.17,205.00)(-6.000,26.503){2}{\rule{0.400pt}{1.083pt}}
\multiput(1424.93,236.00)(-0.489,1.776){15}{\rule{0.118pt}{1.478pt}}
\multiput(1425.17,236.00)(-9.000,27.933){2}{\rule{0.400pt}{0.739pt}}
\multiput(1415.92,267.00)(-0.492,1.439){19}{\rule{0.118pt}{1.227pt}}
\multiput(1416.17,267.00)(-11.000,28.453){2}{\rule{0.400pt}{0.614pt}}
\multiput(1404.92,298.00)(-0.495,0.949){31}{\rule{0.119pt}{0.853pt}}
\multiput(1405.17,298.00)(-17.000,30.230){2}{\rule{0.400pt}{0.426pt}}
\multiput(1387.92,330.00)(-0.497,0.661){47}{\rule{0.120pt}{0.628pt}}
\multiput(1388.17,330.00)(-25.000,31.697){2}{\rule{0.400pt}{0.314pt}}
\multiput(1361.15,363.58)(-0.735,0.497){57}{\rule{0.687pt}{0.120pt}}
\multiput(1362.57,362.17)(-42.575,30.000){2}{\rule{0.343pt}{0.400pt}}
\multiput(1316.50,393.58)(-0.933,0.497){53}{\rule{0.843pt}{0.120pt}}
\multiput(1318.25,392.17)(-50.251,28.000){2}{\rule{0.421pt}{0.400pt}}
\put(1440,113){\circle{12}}
\put(1439,144){\circle{12}}
\put(1437,175){\circle{12}}
\put(1432,205){\circle{12}}
\put(1426,236){\circle{12}}
\put(1417,267){\circle{12}}
\put(1406,298){\circle{12}}
\put(1389,330){\circle{12}}
\put(1364,363){\circle{12}}
\put(1320,393){\circle{12}}
\put(1268,421){\circle{12}}
\put(529,113){\usebox{\plotpoint}}
\put(527.67,113){\rule{0.400pt}{7.468pt}}
\multiput(528.17,113.00)(-1.000,15.500){2}{\rule{0.400pt}{3.734pt}}
\put(526.17,144){\rule{0.400pt}{6.300pt}}
\multiput(527.17,144.00)(-2.000,17.924){2}{\rule{0.400pt}{3.150pt}}
\multiput(524.94,175.00)(-0.468,4.283){5}{\rule{0.113pt}{3.100pt}}
\multiput(525.17,175.00)(-4.000,23.566){2}{\rule{0.400pt}{1.550pt}}
\multiput(520.93,205.00)(-0.477,3.382){7}{\rule{0.115pt}{2.580pt}}
\multiput(521.17,205.00)(-5.000,25.645){2}{\rule{0.400pt}{1.290pt}}
\multiput(515.93,236.00)(-0.485,2.323){11}{\rule{0.117pt}{1.871pt}}
\multiput(516.17,236.00)(-7.000,27.116){2}{\rule{0.400pt}{0.936pt}}
\multiput(508.93,267.00)(-0.488,2.013){13}{\rule{0.117pt}{1.650pt}}
\multiput(509.17,267.00)(-8.000,27.575){2}{\rule{0.400pt}{0.825pt}}
\multiput(500.92,298.00)(-0.491,1.538){17}{\rule{0.118pt}{1.300pt}}
\multiput(501.17,298.00)(-10.000,27.302){2}{\rule{0.400pt}{0.650pt}}
\multiput(490.92,328.00)(-0.492,1.439){19}{\rule{0.118pt}{1.227pt}}
\multiput(491.17,328.00)(-11.000,28.453){2}{\rule{0.400pt}{0.614pt}}
\multiput(479.92,359.00)(-0.493,1.210){23}{\rule{0.119pt}{1.054pt}}
\multiput(480.17,359.00)(-13.000,28.813){2}{\rule{0.400pt}{0.527pt}}
\multiput(466.92,390.00)(-0.494,1.121){25}{\rule{0.119pt}{0.986pt}}
\multiput(467.17,390.00)(-14.000,28.954){2}{\rule{0.400pt}{0.493pt}}
\multiput(452.92,421.00)(-0.494,1.010){27}{\rule{0.119pt}{0.900pt}}
\multiput(453.17,421.00)(-15.000,28.132){2}{\rule{0.400pt}{0.450pt}}
\multiput(437.92,451.00)(-0.494,0.977){29}{\rule{0.119pt}{0.875pt}}
\multiput(438.17,451.00)(-16.000,29.184){2}{\rule{0.400pt}{0.438pt}}
\multiput(421.92,482.00)(-0.495,0.919){31}{\rule{0.119pt}{0.829pt}}
\multiput(422.17,482.00)(-17.000,29.279){2}{\rule{0.400pt}{0.415pt}}
\multiput(404.92,513.00)(-0.495,0.820){35}{\rule{0.119pt}{0.753pt}}
\multiput(405.17,513.00)(-19.000,29.438){2}{\rule{0.400pt}{0.376pt}}
\multiput(385.92,544.00)(-0.495,0.793){35}{\rule{0.119pt}{0.732pt}}
\multiput(386.17,544.00)(-19.000,28.482){2}{\rule{0.400pt}{0.366pt}}
\multiput(366.92,574.00)(-0.496,0.778){37}{\rule{0.119pt}{0.720pt}}
\multiput(367.17,574.00)(-20.000,29.506){2}{\rule{0.400pt}{0.360pt}}
\multiput(346.92,605.00)(-0.496,0.740){39}{\rule{0.119pt}{0.690pt}}
\multiput(347.17,605.00)(-21.000,29.567){2}{\rule{0.400pt}{0.345pt}}
\multiput(325.92,636.00)(-0.496,0.740){39}{\rule{0.119pt}{0.690pt}}
\multiput(326.17,636.00)(-21.000,29.567){2}{\rule{0.400pt}{0.345pt}}
\multiput(304.92,667.00)(-0.496,0.653){43}{\rule{0.120pt}{0.622pt}}
\multiput(305.17,667.00)(-23.000,28.710){2}{\rule{0.400pt}{0.311pt}}
\multiput(281.92,697.00)(-0.496,0.675){43}{\rule{0.120pt}{0.639pt}}
\multiput(282.17,697.00)(-23.000,29.673){2}{\rule{0.400pt}{0.320pt}}
\multiput(258.92,728.00)(-0.496,0.675){43}{\rule{0.120pt}{0.639pt}}
\multiput(259.17,728.00)(-23.000,29.673){2}{\rule{0.400pt}{0.320pt}}
\multiput(235.92,759.00)(-0.495,0.648){31}{\rule{0.119pt}{0.618pt}}
\multiput(236.17,759.00)(-17.000,20.718){2}{\rule{0.400pt}{0.309pt}}
\put(529,113){\circle{12}}
\put(528,144){\circle{12}}
\put(526,175){\circle{12}}
\put(522,205){\circle{12}}
\put(517,236){\circle{12}}
\put(510,267){\circle{12}}
\put(502,298){\circle{12}}
\put(492,328){\circle{12}}
\put(481,359){\circle{12}}
\put(468,390){\circle{12}}
\put(454,421){\circle{12}}
\put(439,451){\circle{12}}
\put(423,482){\circle{12}}
\put(406,513){\circle{12}}
\put(387,544){\circle{12}}
\put(368,574){\circle{12}}
\put(348,605){\circle{12}}
\put(327,636){\circle{12}}
\put(306,667){\circle{12}}
\put(283,697){\circle{12}}
\put(260,728){\circle{12}}
\put(237,759){\circle{12}}
\put(1690,361.67){\rule{29.149pt}{0.400pt}}
\multiput(1750.50,361.17)(-60.500,1.000){2}{\rule{14.574pt}{0.400pt}}
\multiput(1624.00,363.60)(-22.999,0.468){5}{\rule{15.900pt}{0.113pt}}
\multiput(1657.00,362.17)(-124.999,4.000){2}{\rule{7.950pt}{0.400pt}}
\multiput(1505.43,367.59)(-8.286,0.488){13}{\rule{6.400pt}{0.117pt}}
\multiput(1518.72,366.17)(-112.716,8.000){2}{\rule{3.200pt}{0.400pt}}
\multiput(1398.24,375.58)(-2.242,0.495){35}{\rule{1.868pt}{0.119pt}}
\multiput(1402.12,374.17)(-80.122,19.000){2}{\rule{0.934pt}{0.400pt}}
\multiput(1318.26,394.58)(-1.005,0.497){51}{\rule{0.900pt}{0.120pt}}
\multiput(1320.13,393.17)(-52.132,27.000){2}{\rule{0.450pt}{0.400pt}}
\put(1690,363){\circle{12}}
\put(1532,367){\circle{12}}
\put(1406,375){\circle{12}}
\put(1322,394){\circle{12}}
\put(1268,421){\circle{12}}
\sbox{\plotpoint}{\rule[-0.600pt]{1.200pt}{1.200pt}}%
\put(535,1006.51){\rule{1.204pt}{1.200pt}}
\multiput(535.00,1008.51)(2.500,-4.000){2}{\rule{0.602pt}{1.200pt}}
\multiput(542.24,1000.15)(0.503,-0.430){6}{\rule{0.121pt}{1.650pt}}
\multiput(537.51,1003.58)(8.000,-5.575){2}{\rule{1.200pt}{0.825pt}}
\multiput(548.00,995.26)(0.396,-0.502){8}{\rule{1.500pt}{0.121pt}}
\multiput(548.00,995.51)(5.887,-9.000){2}{\rule{0.750pt}{1.200pt}}
\multiput(557.00,986.26)(0.430,-0.503){6}{\rule{1.650pt}{0.121pt}}
\multiput(557.00,986.51)(5.575,-8.000){2}{\rule{0.825pt}{1.200pt}}
\multiput(566.00,978.26)(0.396,-0.502){8}{\rule{1.500pt}{0.121pt}}
\multiput(566.00,978.51)(5.887,-9.000){2}{\rule{0.750pt}{1.200pt}}
\multiput(577.24,965.15)(0.503,-0.430){6}{\rule{0.121pt}{1.650pt}}
\multiput(572.51,968.58)(8.000,-5.575){2}{\rule{1.200pt}{0.825pt}}
\multiput(583.00,960.26)(0.430,-0.503){6}{\rule{1.650pt}{0.121pt}}
\multiput(583.00,960.51)(5.575,-8.000){2}{\rule{0.825pt}{1.200pt}}
\multiput(592.00,952.26)(0.396,-0.502){8}{\rule{1.500pt}{0.121pt}}
\multiput(592.00,952.51)(5.887,-9.000){2}{\rule{0.750pt}{1.200pt}}
\multiput(601.00,943.26)(0.396,-0.502){8}{\rule{1.500pt}{0.121pt}}
\multiput(601.00,943.51)(5.887,-9.000){2}{\rule{0.750pt}{1.200pt}}
\multiput(610.00,934.26)(0.430,-0.503){6}{\rule{1.650pt}{0.121pt}}
\multiput(610.00,934.51)(5.575,-8.000){2}{\rule{0.825pt}{1.200pt}}
\multiput(621.24,922.15)(0.503,-0.430){6}{\rule{0.121pt}{1.650pt}}
\multiput(616.51,925.58)(8.000,-5.575){2}{\rule{1.200pt}{0.825pt}}
\multiput(627.00,917.26)(0.430,-0.503){6}{\rule{1.650pt}{0.121pt}}
\multiput(627.00,917.51)(5.575,-8.000){2}{\rule{0.825pt}{1.200pt}}
\multiput(636.00,909.26)(0.396,-0.502){8}{\rule{1.500pt}{0.121pt}}
\multiput(636.00,909.51)(5.887,-9.000){2}{\rule{0.750pt}{1.200pt}}
\multiput(645.00,900.26)(0.430,-0.503){6}{\rule{1.650pt}{0.121pt}}
\multiput(645.00,900.51)(5.575,-8.000){2}{\rule{0.825pt}{1.200pt}}
\multiput(654.00,892.26)(0.355,-0.503){6}{\rule{1.500pt}{0.121pt}}
\multiput(654.00,892.51)(4.887,-8.000){2}{\rule{0.750pt}{1.200pt}}
\multiput(662.00,884.26)(0.396,-0.502){8}{\rule{1.500pt}{0.121pt}}
\multiput(662.00,884.51)(5.887,-9.000){2}{\rule{0.750pt}{1.200pt}}
\multiput(671.00,875.26)(0.430,-0.503){6}{\rule{1.650pt}{0.121pt}}
\multiput(671.00,875.51)(5.575,-8.000){2}{\rule{0.825pt}{1.200pt}}
\multiput(680.00,867.26)(0.430,-0.503){6}{\rule{1.650pt}{0.121pt}}
\multiput(680.00,867.51)(5.575,-8.000){2}{\rule{0.825pt}{1.200pt}}
\multiput(689.00,859.26)(0.396,-0.502){8}{\rule{1.500pt}{0.121pt}}
\multiput(689.00,859.51)(5.887,-9.000){2}{\rule{0.750pt}{1.200pt}}
\multiput(698.00,850.26)(0.355,-0.503){6}{\rule{1.500pt}{0.121pt}}
\multiput(698.00,850.51)(4.887,-8.000){2}{\rule{0.750pt}{1.200pt}}
\multiput(706.00,842.26)(0.430,-0.503){6}{\rule{1.650pt}{0.121pt}}
\multiput(706.00,842.51)(5.575,-8.000){2}{\rule{0.825pt}{1.200pt}}
\multiput(715.00,834.26)(0.430,-0.503){6}{\rule{1.650pt}{0.121pt}}
\multiput(715.00,834.51)(5.575,-8.000){2}{\rule{0.825pt}{1.200pt}}
\multiput(724.00,826.26)(0.430,-0.503){6}{\rule{1.650pt}{0.121pt}}
\multiput(724.00,826.51)(5.575,-8.000){2}{\rule{0.825pt}{1.200pt}}
\multiput(733.00,818.26)(0.396,-0.502){8}{\rule{1.500pt}{0.121pt}}
\multiput(733.00,818.51)(5.887,-9.000){2}{\rule{0.750pt}{1.200pt}}
\multiput(742.00,809.26)(0.355,-0.503){6}{\rule{1.500pt}{0.121pt}}
\multiput(742.00,809.51)(4.887,-8.000){2}{\rule{0.750pt}{1.200pt}}
\multiput(750.00,801.26)(0.430,-0.503){6}{\rule{1.650pt}{0.121pt}}
\multiput(750.00,801.51)(5.575,-8.000){2}{\rule{0.825pt}{1.200pt}}
\multiput(759.00,793.26)(0.430,-0.503){6}{\rule{1.650pt}{0.121pt}}
\multiput(759.00,793.51)(5.575,-8.000){2}{\rule{0.825pt}{1.200pt}}
\multiput(768.00,785.26)(0.430,-0.503){6}{\rule{1.650pt}{0.121pt}}
\multiput(768.00,785.51)(5.575,-8.000){2}{\rule{0.825pt}{1.200pt}}
\multiput(777.00,777.26)(0.354,-0.505){4}{\rule{1.671pt}{0.122pt}}
\multiput(777.00,777.51)(4.531,-7.000){2}{\rule{0.836pt}{1.200pt}}
\multiput(785.00,770.26)(0.430,-0.503){6}{\rule{1.650pt}{0.121pt}}
\multiput(785.00,770.51)(5.575,-8.000){2}{\rule{0.825pt}{1.200pt}}
\multiput(794.00,762.26)(0.430,-0.503){6}{\rule{1.650pt}{0.121pt}}
\multiput(794.00,762.51)(5.575,-8.000){2}{\rule{0.825pt}{1.200pt}}
\multiput(803.00,754.26)(0.430,-0.503){6}{\rule{1.650pt}{0.121pt}}
\multiput(803.00,754.51)(5.575,-8.000){2}{\rule{0.825pt}{1.200pt}}
\multiput(812.00,746.26)(0.430,-0.503){6}{\rule{1.650pt}{0.121pt}}
\multiput(812.00,746.51)(5.575,-8.000){2}{\rule{0.825pt}{1.200pt}}
\multiput(821.00,738.26)(0.354,-0.505){4}{\rule{1.671pt}{0.122pt}}
\multiput(821.00,738.51)(4.531,-7.000){2}{\rule{0.836pt}{1.200pt}}
\multiput(829.00,731.26)(0.430,-0.503){6}{\rule{1.650pt}{0.121pt}}
\multiput(829.00,731.51)(5.575,-8.000){2}{\rule{0.825pt}{1.200pt}}
\multiput(838.00,723.26)(0.450,-0.505){4}{\rule{1.843pt}{0.122pt}}
\multiput(838.00,723.51)(5.175,-7.000){2}{\rule{0.921pt}{1.200pt}}
\multiput(847.00,716.26)(0.430,-0.503){6}{\rule{1.650pt}{0.121pt}}
\multiput(847.00,716.51)(5.575,-8.000){2}{\rule{0.825pt}{1.200pt}}
\multiput(856.00,708.26)(0.354,-0.505){4}{\rule{1.671pt}{0.122pt}}
\multiput(856.00,708.51)(4.531,-7.000){2}{\rule{0.836pt}{1.200pt}}
\multiput(864.00,701.26)(0.430,-0.503){6}{\rule{1.650pt}{0.121pt}}
\multiput(864.00,701.51)(5.575,-8.000){2}{\rule{0.825pt}{1.200pt}}
\multiput(873.00,693.26)(0.450,-0.505){4}{\rule{1.843pt}{0.122pt}}
\multiput(873.00,693.51)(5.175,-7.000){2}{\rule{0.921pt}{1.200pt}}
\multiput(882.00,686.26)(0.430,-0.503){6}{\rule{1.650pt}{0.121pt}}
\multiput(882.00,686.51)(5.575,-8.000){2}{\rule{0.825pt}{1.200pt}}
\multiput(891.00,678.26)(0.450,-0.505){4}{\rule{1.843pt}{0.122pt}}
\multiput(891.00,678.51)(5.175,-7.000){2}{\rule{0.921pt}{1.200pt}}
\multiput(900.00,671.26)(0.354,-0.505){4}{\rule{1.671pt}{0.122pt}}
\multiput(900.00,671.51)(4.531,-7.000){2}{\rule{0.836pt}{1.200pt}}
\multiput(908.00,664.26)(0.450,-0.505){4}{\rule{1.843pt}{0.122pt}}
\multiput(908.00,664.51)(5.175,-7.000){2}{\rule{0.921pt}{1.200pt}}
\multiput(917.00,657.26)(0.450,-0.505){4}{\rule{1.843pt}{0.122pt}}
\multiput(917.00,657.51)(5.175,-7.000){2}{\rule{0.921pt}{1.200pt}}
\multiput(926.00,650.26)(0.450,-0.505){4}{\rule{1.843pt}{0.122pt}}
\multiput(926.00,650.51)(5.175,-7.000){2}{\rule{0.921pt}{1.200pt}}
\multiput(935.00,643.26)(0.354,-0.505){4}{\rule{1.671pt}{0.122pt}}
\multiput(935.00,643.51)(4.531,-7.000){2}{\rule{0.836pt}{1.200pt}}
\multiput(943.00,636.26)(0.450,-0.505){4}{\rule{1.843pt}{0.122pt}}
\multiput(943.00,636.51)(5.175,-7.000){2}{\rule{0.921pt}{1.200pt}}
\multiput(952.00,629.26)(0.450,-0.505){4}{\rule{1.843pt}{0.122pt}}
\multiput(952.00,629.51)(5.175,-7.000){2}{\rule{0.921pt}{1.200pt}}
\multiput(961.00,622.26)(0.450,-0.505){4}{\rule{1.843pt}{0.122pt}}
\multiput(961.00,622.51)(5.175,-7.000){2}{\rule{0.921pt}{1.200pt}}
\multiput(970.00,615.26)(0.450,-0.505){4}{\rule{1.843pt}{0.122pt}}
\multiput(970.00,615.51)(5.175,-7.000){2}{\rule{0.921pt}{1.200pt}}
\multiput(979.00,608.26)(0.354,-0.505){4}{\rule{1.671pt}{0.122pt}}
\multiput(979.00,608.51)(4.531,-7.000){2}{\rule{0.836pt}{1.200pt}}
\multiput(987.00,601.25)(0.283,-0.509){2}{\rule{2.100pt}{0.123pt}}
\multiput(987.00,601.51)(4.641,-6.000){2}{\rule{1.050pt}{1.200pt}}
\multiput(996.00,595.26)(0.450,-0.505){4}{\rule{1.843pt}{0.122pt}}
\multiput(996.00,595.51)(5.175,-7.000){2}{\rule{0.921pt}{1.200pt}}
\multiput(1005.00,588.26)(0.450,-0.505){4}{\rule{1.843pt}{0.122pt}}
\multiput(1005.00,588.51)(5.175,-7.000){2}{\rule{0.921pt}{1.200pt}}
\multiput(1014.00,581.25)(0.113,-0.509){2}{\rule{1.900pt}{0.123pt}}
\multiput(1014.00,581.51)(4.056,-6.000){2}{\rule{0.950pt}{1.200pt}}
\multiput(1022.00,575.26)(0.450,-0.505){4}{\rule{1.843pt}{0.122pt}}
\multiput(1022.00,575.51)(5.175,-7.000){2}{\rule{0.921pt}{1.200pt}}
\multiput(1031.00,568.25)(0.283,-0.509){2}{\rule{2.100pt}{0.123pt}}
\multiput(1031.00,568.51)(4.641,-6.000){2}{\rule{1.050pt}{1.200pt}}
\multiput(1040.00,562.25)(0.283,-0.509){2}{\rule{2.100pt}{0.123pt}}
\multiput(1040.00,562.51)(4.641,-6.000){2}{\rule{1.050pt}{1.200pt}}
\multiput(1049.00,556.26)(0.450,-0.505){4}{\rule{1.843pt}{0.122pt}}
\multiput(1049.00,556.51)(5.175,-7.000){2}{\rule{0.921pt}{1.200pt}}
\multiput(1058.00,549.25)(0.113,-0.509){2}{\rule{1.900pt}{0.123pt}}
\multiput(1058.00,549.51)(4.056,-6.000){2}{\rule{0.950pt}{1.200pt}}
\multiput(1066.00,543.25)(0.283,-0.509){2}{\rule{2.100pt}{0.123pt}}
\multiput(1066.00,543.51)(4.641,-6.000){2}{\rule{1.050pt}{1.200pt}}
\multiput(1075.00,537.25)(0.283,-0.509){2}{\rule{2.100pt}{0.123pt}}
\multiput(1075.00,537.51)(4.641,-6.000){2}{\rule{1.050pt}{1.200pt}}
\multiput(1084.00,531.25)(0.283,-0.509){2}{\rule{2.100pt}{0.123pt}}
\multiput(1084.00,531.51)(4.641,-6.000){2}{\rule{1.050pt}{1.200pt}}
\multiput(1093.00,525.25)(0.113,-0.509){2}{\rule{1.900pt}{0.123pt}}
\multiput(1093.00,525.51)(4.056,-6.000){2}{\rule{0.950pt}{1.200pt}}
\multiput(1101.00,519.25)(0.283,-0.509){2}{\rule{2.100pt}{0.123pt}}
\multiput(1101.00,519.51)(4.641,-6.000){2}{\rule{1.050pt}{1.200pt}}
\multiput(1110.00,513.25)(0.283,-0.509){2}{\rule{2.100pt}{0.123pt}}
\multiput(1110.00,513.51)(4.641,-6.000){2}{\rule{1.050pt}{1.200pt}}
\put(1119,505.01){\rule{2.168pt}{1.200pt}}
\multiput(1119.00,507.51)(4.500,-5.000){2}{\rule{1.084pt}{1.200pt}}
\multiput(1128.00,502.25)(0.283,-0.509){2}{\rule{2.100pt}{0.123pt}}
\multiput(1128.00,502.51)(4.641,-6.000){2}{\rule{1.050pt}{1.200pt}}
\multiput(1137.00,496.25)(0.113,-0.509){2}{\rule{1.900pt}{0.123pt}}
\multiput(1137.00,496.51)(4.056,-6.000){2}{\rule{0.950pt}{1.200pt}}
\put(1145,488.01){\rule{2.168pt}{1.200pt}}
\multiput(1145.00,490.51)(4.500,-5.000){2}{\rule{1.084pt}{1.200pt}}
\multiput(1154.00,485.25)(0.283,-0.509){2}{\rule{2.100pt}{0.123pt}}
\multiput(1154.00,485.51)(4.641,-6.000){2}{\rule{1.050pt}{1.200pt}}
\put(1163,477.01){\rule{2.168pt}{1.200pt}}
\multiput(1163.00,479.51)(4.500,-5.000){2}{\rule{1.084pt}{1.200pt}}
\multiput(1172.00,474.25)(0.113,-0.509){2}{\rule{1.900pt}{0.123pt}}
\multiput(1172.00,474.51)(4.056,-6.000){2}{\rule{0.950pt}{1.200pt}}
\put(1180,466.01){\rule{2.168pt}{1.200pt}}
\multiput(1180.00,468.51)(4.500,-5.000){2}{\rule{1.084pt}{1.200pt}}
\put(1189,461.01){\rule{2.168pt}{1.200pt}}
\multiput(1189.00,463.51)(4.500,-5.000){2}{\rule{1.084pt}{1.200pt}}
\multiput(1198.00,458.25)(0.283,-0.509){2}{\rule{2.100pt}{0.123pt}}
\multiput(1198.00,458.51)(4.641,-6.000){2}{\rule{1.050pt}{1.200pt}}
\put(1207,450.01){\rule{2.168pt}{1.200pt}}
\multiput(1207.00,452.51)(4.500,-5.000){2}{\rule{1.084pt}{1.200pt}}
\put(1216,445.01){\rule{1.927pt}{1.200pt}}
\multiput(1216.00,447.51)(4.000,-5.000){2}{\rule{0.964pt}{1.200pt}}
\put(1224,440.01){\rule{2.168pt}{1.200pt}}
\multiput(1224.00,442.51)(4.500,-5.000){2}{\rule{1.084pt}{1.200pt}}
\put(1233,435.01){\rule{2.168pt}{1.200pt}}
\multiput(1233.00,437.51)(4.500,-5.000){2}{\rule{1.084pt}{1.200pt}}
\put(1242,430.01){\rule{2.168pt}{1.200pt}}
\multiput(1242.00,432.51)(4.500,-5.000){2}{\rule{1.084pt}{1.200pt}}
\put(1251,425.01){\rule{1.927pt}{1.200pt}}
\multiput(1251.00,427.51)(4.000,-5.000){2}{\rule{0.964pt}{1.200pt}}
\put(1259,420.51){\rule{2.168pt}{1.200pt}}
\multiput(1259.00,422.51)(4.500,-4.000){2}{\rule{1.084pt}{1.200pt}}
\sbox{\plotpoint}{\rule[-0.200pt]{0.400pt}{0.400pt}}%
\put(1371,362){\circle*{12}}
\put(1420,292){\circle*{12}}
\put(1433,225){\circle*{12}}
\put(1404,371){\circle*{12}}
\put(1709,357){\circle*{12}}
\put(1300,396){\raisebox{-.8pt}{\makebox(0,0){$\Diamond$}}}
\put(1317,377){\raisebox{-.8pt}{\makebox(0,0){$\Diamond$}}}
\put(1358,340){\raisebox{-.8pt}{\makebox(0,0){$\Diamond$}}}
\put(1402,287){\raisebox{-.8pt}{\makebox(0,0){$\Diamond$}}}
\put(1421,245){\raisebox{-.8pt}{\makebox(0,0){$\Diamond$}}}
\put(1428,211){\raisebox{-.8pt}{\makebox(0,0){$\Diamond$}}}
\put(261,734){\raisebox{-.8pt}{\makebox(0,0){$\Diamond$}}}
\put(335,630){\raisebox{-.8pt}{\makebox(0,0){$\Diamond$}}}
\put(416,497){\raisebox{-.8pt}{\makebox(0,0){$\Diamond$}}}
\put(452,426){\raisebox{-.8pt}{\makebox(0,0){$\Diamond$}}}
\put(499,307){\raisebox{-.8pt}{\makebox(0,0){$\Diamond$}}}
\put(510,266){\raisebox{-.8pt}{\makebox(0,0){$\Diamond$}}}
\put(529,113){\raisebox{-.8pt}{\makebox(0,0){$\Diamond$}}}
\sbox{\plotpoint}{\rule[-0.600pt]{1.200pt}{1.200pt}}%
\put(535,1006.51){\rule{1.204pt}{1.200pt}}
\multiput(535.00,1008.51)(2.500,-4.000){2}{\rule{0.602pt}{1.200pt}}
\multiput(542.24,1000.15)(0.503,-0.430){6}{\rule{0.121pt}{1.650pt}}
\multiput(537.51,1003.58)(8.000,-5.575){2}{\rule{1.200pt}{0.825pt}}
\multiput(548.00,995.26)(0.396,-0.502){8}{\rule{1.500pt}{0.121pt}}
\multiput(548.00,995.51)(5.887,-9.000){2}{\rule{0.750pt}{1.200pt}}
\multiput(557.00,986.26)(0.430,-0.503){6}{\rule{1.650pt}{0.121pt}}
\multiput(557.00,986.51)(5.575,-8.000){2}{\rule{0.825pt}{1.200pt}}
\multiput(566.00,978.26)(0.396,-0.502){8}{\rule{1.500pt}{0.121pt}}
\multiput(566.00,978.51)(5.887,-9.000){2}{\rule{0.750pt}{1.200pt}}
\multiput(577.24,965.15)(0.503,-0.430){6}{\rule{0.121pt}{1.650pt}}
\multiput(572.51,968.58)(8.000,-5.575){2}{\rule{1.200pt}{0.825pt}}
\multiput(583.00,960.26)(0.430,-0.503){6}{\rule{1.650pt}{0.121pt}}
\multiput(583.00,960.51)(5.575,-8.000){2}{\rule{0.825pt}{1.200pt}}
\multiput(592.00,952.26)(0.396,-0.502){8}{\rule{1.500pt}{0.121pt}}
\multiput(592.00,952.51)(5.887,-9.000){2}{\rule{0.750pt}{1.200pt}}
\multiput(601.00,943.26)(0.396,-0.502){8}{\rule{1.500pt}{0.121pt}}
\multiput(601.00,943.51)(5.887,-9.000){2}{\rule{0.750pt}{1.200pt}}
\multiput(610.00,934.26)(0.430,-0.503){6}{\rule{1.650pt}{0.121pt}}
\multiput(610.00,934.51)(5.575,-8.000){2}{\rule{0.825pt}{1.200pt}}
\multiput(621.24,922.15)(0.503,-0.430){6}{\rule{0.121pt}{1.650pt}}
\multiput(616.51,925.58)(8.000,-5.575){2}{\rule{1.200pt}{0.825pt}}
\multiput(627.00,917.26)(0.430,-0.503){6}{\rule{1.650pt}{0.121pt}}
\multiput(627.00,917.51)(5.575,-8.000){2}{\rule{0.825pt}{1.200pt}}
\multiput(636.00,909.26)(0.396,-0.502){8}{\rule{1.500pt}{0.121pt}}
\multiput(636.00,909.51)(5.887,-9.000){2}{\rule{0.750pt}{1.200pt}}
\multiput(645.00,900.26)(0.430,-0.503){6}{\rule{1.650pt}{0.121pt}}
\multiput(645.00,900.51)(5.575,-8.000){2}{\rule{0.825pt}{1.200pt}}
\multiput(654.00,892.26)(0.355,-0.503){6}{\rule{1.500pt}{0.121pt}}
\multiput(654.00,892.51)(4.887,-8.000){2}{\rule{0.750pt}{1.200pt}}
\multiput(662.00,884.26)(0.396,-0.502){8}{\rule{1.500pt}{0.121pt}}
\multiput(662.00,884.51)(5.887,-9.000){2}{\rule{0.750pt}{1.200pt}}
\multiput(671.00,875.26)(0.430,-0.503){6}{\rule{1.650pt}{0.121pt}}
\multiput(671.00,875.51)(5.575,-8.000){2}{\rule{0.825pt}{1.200pt}}
\multiput(680.00,867.26)(0.430,-0.503){6}{\rule{1.650pt}{0.121pt}}
\multiput(680.00,867.51)(5.575,-8.000){2}{\rule{0.825pt}{1.200pt}}
\multiput(689.00,859.26)(0.396,-0.502){8}{\rule{1.500pt}{0.121pt}}
\multiput(689.00,859.51)(5.887,-9.000){2}{\rule{0.750pt}{1.200pt}}
\multiput(698.00,850.26)(0.355,-0.503){6}{\rule{1.500pt}{0.121pt}}
\multiput(698.00,850.51)(4.887,-8.000){2}{\rule{0.750pt}{1.200pt}}
\multiput(706.00,842.26)(0.430,-0.503){6}{\rule{1.650pt}{0.121pt}}
\multiput(706.00,842.51)(5.575,-8.000){2}{\rule{0.825pt}{1.200pt}}
\multiput(715.00,834.26)(0.430,-0.503){6}{\rule{1.650pt}{0.121pt}}
\multiput(715.00,834.51)(5.575,-8.000){2}{\rule{0.825pt}{1.200pt}}
\multiput(724.00,826.26)(0.430,-0.503){6}{\rule{1.650pt}{0.121pt}}
\multiput(724.00,826.51)(5.575,-8.000){2}{\rule{0.825pt}{1.200pt}}
\multiput(733.00,818.26)(0.396,-0.502){8}{\rule{1.500pt}{0.121pt}}
\multiput(733.00,818.51)(5.887,-9.000){2}{\rule{0.750pt}{1.200pt}}
\multiput(742.00,809.26)(0.355,-0.503){6}{\rule{1.500pt}{0.121pt}}
\multiput(742.00,809.51)(4.887,-8.000){2}{\rule{0.750pt}{1.200pt}}
\multiput(750.00,801.26)(0.430,-0.503){6}{\rule{1.650pt}{0.121pt}}
\multiput(750.00,801.51)(5.575,-8.000){2}{\rule{0.825pt}{1.200pt}}
\multiput(759.00,793.26)(0.430,-0.503){6}{\rule{1.650pt}{0.121pt}}
\multiput(759.00,793.51)(5.575,-8.000){2}{\rule{0.825pt}{1.200pt}}
\multiput(768.00,785.26)(0.430,-0.503){6}{\rule{1.650pt}{0.121pt}}
\multiput(768.00,785.51)(5.575,-8.000){2}{\rule{0.825pt}{1.200pt}}
\multiput(777.00,777.26)(0.354,-0.505){4}{\rule{1.671pt}{0.122pt}}
\multiput(777.00,777.51)(4.531,-7.000){2}{\rule{0.836pt}{1.200pt}}
\multiput(785.00,770.26)(0.430,-0.503){6}{\rule{1.650pt}{0.121pt}}
\multiput(785.00,770.51)(5.575,-8.000){2}{\rule{0.825pt}{1.200pt}}
\multiput(794.00,762.26)(0.430,-0.503){6}{\rule{1.650pt}{0.121pt}}
\multiput(794.00,762.51)(5.575,-8.000){2}{\rule{0.825pt}{1.200pt}}
\multiput(803.00,754.26)(0.430,-0.503){6}{\rule{1.650pt}{0.121pt}}
\multiput(803.00,754.51)(5.575,-8.000){2}{\rule{0.825pt}{1.200pt}}
\multiput(812.00,746.26)(0.430,-0.503){6}{\rule{1.650pt}{0.121pt}}
\multiput(812.00,746.51)(5.575,-8.000){2}{\rule{0.825pt}{1.200pt}}
\multiput(821.00,738.26)(0.354,-0.505){4}{\rule{1.671pt}{0.122pt}}
\multiput(821.00,738.51)(4.531,-7.000){2}{\rule{0.836pt}{1.200pt}}
\multiput(829.00,731.26)(0.430,-0.503){6}{\rule{1.650pt}{0.121pt}}
\multiput(829.00,731.51)(5.575,-8.000){2}{\rule{0.825pt}{1.200pt}}
\multiput(838.00,723.26)(0.450,-0.505){4}{\rule{1.843pt}{0.122pt}}
\multiput(838.00,723.51)(5.175,-7.000){2}{\rule{0.921pt}{1.200pt}}
\multiput(847.00,716.26)(0.430,-0.503){6}{\rule{1.650pt}{0.121pt}}
\multiput(847.00,716.51)(5.575,-8.000){2}{\rule{0.825pt}{1.200pt}}
\multiput(856.00,708.26)(0.354,-0.505){4}{\rule{1.671pt}{0.122pt}}
\multiput(856.00,708.51)(4.531,-7.000){2}{\rule{0.836pt}{1.200pt}}
\multiput(864.00,701.26)(0.430,-0.503){6}{\rule{1.650pt}{0.121pt}}
\multiput(864.00,701.51)(5.575,-8.000){2}{\rule{0.825pt}{1.200pt}}
\multiput(873.00,693.26)(0.450,-0.505){4}{\rule{1.843pt}{0.122pt}}
\multiput(873.00,693.51)(5.175,-7.000){2}{\rule{0.921pt}{1.200pt}}
\multiput(882.00,686.26)(0.430,-0.503){6}{\rule{1.650pt}{0.121pt}}
\multiput(882.00,686.51)(5.575,-8.000){2}{\rule{0.825pt}{1.200pt}}
\multiput(891.00,678.26)(0.450,-0.505){4}{\rule{1.843pt}{0.122pt}}
\multiput(891.00,678.51)(5.175,-7.000){2}{\rule{0.921pt}{1.200pt}}
\multiput(900.00,671.26)(0.354,-0.505){4}{\rule{1.671pt}{0.122pt}}
\multiput(900.00,671.51)(4.531,-7.000){2}{\rule{0.836pt}{1.200pt}}
\multiput(908.00,664.26)(0.450,-0.505){4}{\rule{1.843pt}{0.122pt}}
\multiput(908.00,664.51)(5.175,-7.000){2}{\rule{0.921pt}{1.200pt}}
\multiput(917.00,657.26)(0.450,-0.505){4}{\rule{1.843pt}{0.122pt}}
\multiput(917.00,657.51)(5.175,-7.000){2}{\rule{0.921pt}{1.200pt}}
\multiput(926.00,650.26)(0.450,-0.505){4}{\rule{1.843pt}{0.122pt}}
\multiput(926.00,650.51)(5.175,-7.000){2}{\rule{0.921pt}{1.200pt}}
\multiput(935.00,643.26)(0.354,-0.505){4}{\rule{1.671pt}{0.122pt}}
\multiput(935.00,643.51)(4.531,-7.000){2}{\rule{0.836pt}{1.200pt}}
\multiput(943.00,636.26)(0.450,-0.505){4}{\rule{1.843pt}{0.122pt}}
\multiput(943.00,636.51)(5.175,-7.000){2}{\rule{0.921pt}{1.200pt}}
\multiput(952.00,629.26)(0.450,-0.505){4}{\rule{1.843pt}{0.122pt}}
\multiput(952.00,629.51)(5.175,-7.000){2}{\rule{0.921pt}{1.200pt}}
\multiput(961.00,622.26)(0.450,-0.505){4}{\rule{1.843pt}{0.122pt}}
\multiput(961.00,622.51)(5.175,-7.000){2}{\rule{0.921pt}{1.200pt}}
\multiput(970.00,615.26)(0.450,-0.505){4}{\rule{1.843pt}{0.122pt}}
\multiput(970.00,615.51)(5.175,-7.000){2}{\rule{0.921pt}{1.200pt}}
\multiput(979.00,608.26)(0.354,-0.505){4}{\rule{1.671pt}{0.122pt}}
\multiput(979.00,608.51)(4.531,-7.000){2}{\rule{0.836pt}{1.200pt}}
\multiput(987.00,601.25)(0.283,-0.509){2}{\rule{2.100pt}{0.123pt}}
\multiput(987.00,601.51)(4.641,-6.000){2}{\rule{1.050pt}{1.200pt}}
\multiput(996.00,595.26)(0.450,-0.505){4}{\rule{1.843pt}{0.122pt}}
\multiput(996.00,595.51)(5.175,-7.000){2}{\rule{0.921pt}{1.200pt}}
\multiput(1005.00,588.26)(0.450,-0.505){4}{\rule{1.843pt}{0.122pt}}
\multiput(1005.00,588.51)(5.175,-7.000){2}{\rule{0.921pt}{1.200pt}}
\multiput(1014.00,581.25)(0.113,-0.509){2}{\rule{1.900pt}{0.123pt}}
\multiput(1014.00,581.51)(4.056,-6.000){2}{\rule{0.950pt}{1.200pt}}
\multiput(1022.00,575.26)(0.450,-0.505){4}{\rule{1.843pt}{0.122pt}}
\multiput(1022.00,575.51)(5.175,-7.000){2}{\rule{0.921pt}{1.200pt}}
\multiput(1031.00,568.25)(0.283,-0.509){2}{\rule{2.100pt}{0.123pt}}
\multiput(1031.00,568.51)(4.641,-6.000){2}{\rule{1.050pt}{1.200pt}}
\multiput(1040.00,562.25)(0.283,-0.509){2}{\rule{2.100pt}{0.123pt}}
\multiput(1040.00,562.51)(4.641,-6.000){2}{\rule{1.050pt}{1.200pt}}
\multiput(1049.00,556.26)(0.450,-0.505){4}{\rule{1.843pt}{0.122pt}}
\multiput(1049.00,556.51)(5.175,-7.000){2}{\rule{0.921pt}{1.200pt}}
\multiput(1058.00,549.25)(0.113,-0.509){2}{\rule{1.900pt}{0.123pt}}
\multiput(1058.00,549.51)(4.056,-6.000){2}{\rule{0.950pt}{1.200pt}}
\multiput(1066.00,543.25)(0.283,-0.509){2}{\rule{2.100pt}{0.123pt}}
\multiput(1066.00,543.51)(4.641,-6.000){2}{\rule{1.050pt}{1.200pt}}
\multiput(1075.00,537.25)(0.283,-0.509){2}{\rule{2.100pt}{0.123pt}}
\multiput(1075.00,537.51)(4.641,-6.000){2}{\rule{1.050pt}{1.200pt}}
\multiput(1084.00,531.25)(0.283,-0.509){2}{\rule{2.100pt}{0.123pt}}
\multiput(1084.00,531.51)(4.641,-6.000){2}{\rule{1.050pt}{1.200pt}}
\multiput(1093.00,525.25)(0.113,-0.509){2}{\rule{1.900pt}{0.123pt}}
\multiput(1093.00,525.51)(4.056,-6.000){2}{\rule{0.950pt}{1.200pt}}
\multiput(1101.00,519.25)(0.283,-0.509){2}{\rule{2.100pt}{0.123pt}}
\multiput(1101.00,519.51)(4.641,-6.000){2}{\rule{1.050pt}{1.200pt}}
\multiput(1110.00,513.25)(0.283,-0.509){2}{\rule{2.100pt}{0.123pt}}
\multiput(1110.00,513.51)(4.641,-6.000){2}{\rule{1.050pt}{1.200pt}}
\put(1119,505.01){\rule{2.168pt}{1.200pt}}
\multiput(1119.00,507.51)(4.500,-5.000){2}{\rule{1.084pt}{1.200pt}}
\multiput(1128.00,502.25)(0.283,-0.509){2}{\rule{2.100pt}{0.123pt}}
\multiput(1128.00,502.51)(4.641,-6.000){2}{\rule{1.050pt}{1.200pt}}
\multiput(1137.00,496.25)(0.113,-0.509){2}{\rule{1.900pt}{0.123pt}}
\multiput(1137.00,496.51)(4.056,-6.000){2}{\rule{0.950pt}{1.200pt}}
\put(1145,488.01){\rule{2.168pt}{1.200pt}}
\multiput(1145.00,490.51)(4.500,-5.000){2}{\rule{1.084pt}{1.200pt}}
\multiput(1154.00,485.25)(0.283,-0.509){2}{\rule{2.100pt}{0.123pt}}
\multiput(1154.00,485.51)(4.641,-6.000){2}{\rule{1.050pt}{1.200pt}}
\put(1163,477.01){\rule{2.168pt}{1.200pt}}
\multiput(1163.00,479.51)(4.500,-5.000){2}{\rule{1.084pt}{1.200pt}}
\multiput(1172.00,474.25)(0.113,-0.509){2}{\rule{1.900pt}{0.123pt}}
\multiput(1172.00,474.51)(4.056,-6.000){2}{\rule{0.950pt}{1.200pt}}
\put(1180,466.01){\rule{2.168pt}{1.200pt}}
\multiput(1180.00,468.51)(4.500,-5.000){2}{\rule{1.084pt}{1.200pt}}
\put(1189,461.01){\rule{2.168pt}{1.200pt}}
\multiput(1189.00,463.51)(4.500,-5.000){2}{\rule{1.084pt}{1.200pt}}
\multiput(1198.00,458.25)(0.283,-0.509){2}{\rule{2.100pt}{0.123pt}}
\multiput(1198.00,458.51)(4.641,-6.000){2}{\rule{1.050pt}{1.200pt}}
\put(1207,450.01){\rule{2.168pt}{1.200pt}}
\multiput(1207.00,452.51)(4.500,-5.000){2}{\rule{1.084pt}{1.200pt}}
\put(1216,445.01){\rule{1.927pt}{1.200pt}}
\multiput(1216.00,447.51)(4.000,-5.000){2}{\rule{0.964pt}{1.200pt}}
\put(1224,440.01){\rule{2.168pt}{1.200pt}}
\multiput(1224.00,442.51)(4.500,-5.000){2}{\rule{1.084pt}{1.200pt}}
\put(1233,435.01){\rule{2.168pt}{1.200pt}}
\multiput(1233.00,437.51)(4.500,-5.000){2}{\rule{1.084pt}{1.200pt}}
\put(1242,430.01){\rule{2.168pt}{1.200pt}}
\multiput(1242.00,432.51)(4.500,-5.000){2}{\rule{1.084pt}{1.200pt}}
\put(1251,425.01){\rule{1.927pt}{1.200pt}}
\multiput(1251.00,427.51)(4.000,-5.000){2}{\rule{0.964pt}{1.200pt}}
\put(1259,420.51){\rule{2.168pt}{1.200pt}}
\multiput(1259.00,422.51)(4.500,-4.000){2}{\rule{1.084pt}{1.200pt}}
\sbox{\plotpoint}{\rule[-0.500pt]{1.000pt}{1.000pt}}%
\put(359,1011){\usebox{\plotpoint}}
\multiput(359,1011)(7.145,-19.487){4}{\usebox{\plotpoint}}
\multiput(381,951)(7.578,-19.323){2}{\usebox{\plotpoint}}
\multiput(401,900)(6.266,-19.787){3}{\usebox{\plotpoint}}
\multiput(420,840)(5.753,-19.942){3}{\usebox{\plotpoint}}
\multiput(435,788)(5.432,-20.032){3}{\usebox{\plotpoint}}
\multiput(451,729)(4.296,-20.306){2}{\usebox{\plotpoint}}
\multiput(462,677)(3.743,-20.415){3}{\usebox{\plotpoint}}
\multiput(473,617)(3.216,-20.505){3}{\usebox{\plotpoint}}
\multiput(481,566)(3.516,-20.456){3}{\usebox{\plotpoint}}
\multiput(492,502)(3.216,-20.505){2}{\usebox{\plotpoint}}
\multiput(500,451)(1.959,-20.663){6}{\usebox{\plotpoint}}
\multiput(511,335)(1.404,-20.708){3}{\usebox{\plotpoint}}
\multiput(515,276)(1.592,-20.694){2}{\usebox{\plotpoint}}
\multiput(519,224)(1.753,-20.681){3}{\usebox{\plotpoint}}
\multiput(524,165)(1.987,-20.660){3}{\usebox{\plotpoint}}
\put(529,113){\usebox{\plotpoint}}
\end{picture}

\end{center}
\caption{Phase diagram of the AT model in two dimensions. The solid bold line
represents the exactly known critical line, which terminates at the
4-state Potts point.
Empty circles with continuous lines describe our results. The solid circles
display MCRG results \protect,  dotted line is drawn
after Baxter and diamonds are the transfer matrix results
\protect combined with conformal invariance.}
\label{res}
\end{figure}
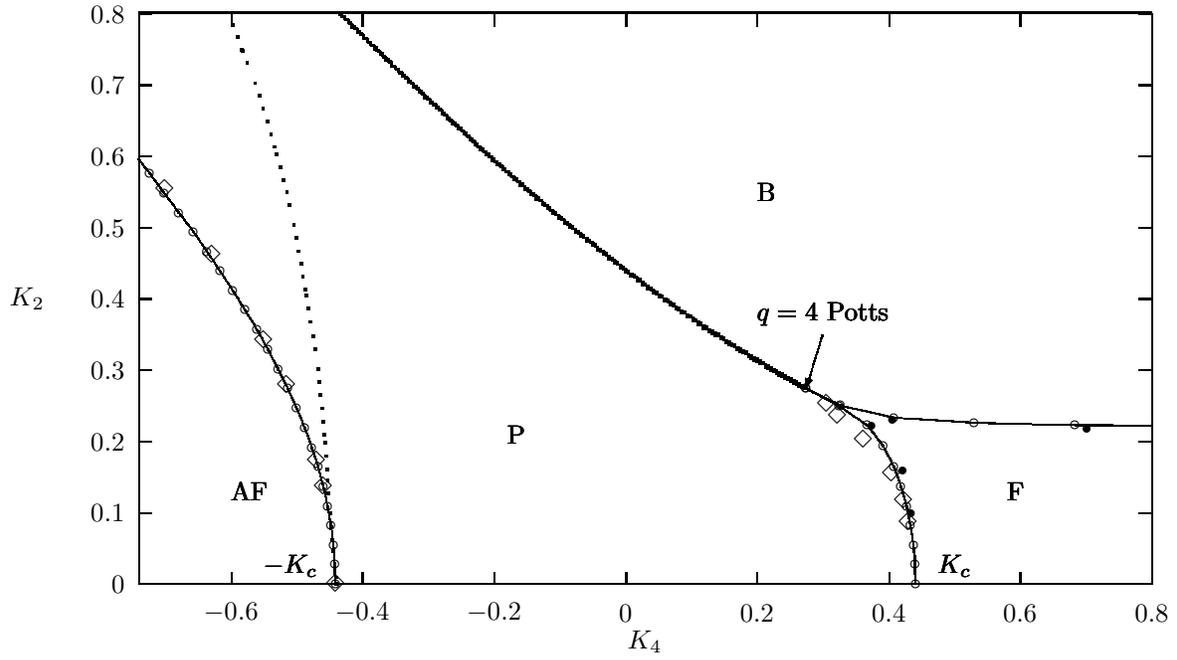

\end{document}